\documentstyle[12pt,epsfig]{article}
\makeatletter
\def\slash#1{{\mathpalette\c@ncel{#1}}} 
\makeatother

\newcommand\beq{\begin{eqnarray}}
\newcommand\eeq{\end{eqnarray}}

\newcommand\la{\langle}
\newcommand\ra{\rangle}

\def\pslash{\rlap/{\mkern-1mu p}}

\def\xhat{\widehat{x}}
\def\zhat{\widehat{z}}

\def\GFt{\widetilde{G}_F}

\begin{document}
\begin{flushright}
\end{flushright}
\vspace*{15mm}
\begin{center}

{\Large \bf Master Formula for Twist-3 Soft-Gluon-Pole\\[4mm] Mechanism
to Single Transverse-Spin Asymmetry
}
\vspace{1.5cm}\\
 {\sc Yuji Koike$^1$ and Kazuhiro Tanaka$^2$}
\\[0.4cm]
\vspace*{0.1cm}{\it $^1$ Department of Physics, Niigata University,
Ikarashi, Niigata 950-2181, Japan}\\
\vspace*{0.1cm}{\it $^2$ Department of Physics, 
Juntendo University, Inba-gun, Chiba 270-1695, Japan}
\\[3cm]

{\large \bf Abstract} \end{center}
Perturbative QCD relates the single transverse-spin asymmetries (SSAs) 
for hard processes 
at large transverse-momentum of produced particle
to partonic matrix elements that
describe interference between scattering from
a coherent quark-gluon pair and from a single quark, generated through 
twist-3 quark-gluon correlations inside a hadron.
When the coherent gluon is soft at the 
gluonic poles,
its coupling to partonic subprocess can be systematically disentangled, so that 
the relevant 
interfering amplitude
can be derived entirely from 
the Born diagrams
for the scattering from a single quark.
We establish a new formula that represents
the exact rules to derive the SSA due to soft-gluon poles
from the knowledge of the twist-2 cross section formula for unpolarized 
processes.  
This single master formula is applicable to 
a range of processes 
like Drell-Yan and direct-photon production,
and semi-inclusive deep inelastic scattering,
and is also useful to manifest the gauge invariance of the results.

\noindent

\newpage

There have been 
two systematic frameworks to study single spin asymmetries (SSAs)
observed in a variety of high energy semi-inclusive reactions.
One is based on the so-called ``$T$-odd'' distribution and fragmentation functions
with parton's intrinsic transverse momentum\,\cite{Sivers,Collins93}.  It describes
SSAs in the region of the small transverse momentum $p_T$ of the 
final hadrons as a leading twist effect.  Its factorization property
and the universality of ``$T$-odd'' functions have been extensively studied\,\cite{JMY05}.   
The other one is the twist-3 mechanism based on the collinear factorization, and
is suited to describe SSAs in the large $p_T$ region\,\cite{ET82,QS91,ekt06}. 
It relates the SSA to certain quark-gluon correlation functions in the hadrons.
Recently it has been shown that these two
mechanisms give the identical SSA in the intermediate $p_T$ region
for Drell-Yan and semi-inclusive deep inelastic scatterings, i.e., 
describe the same physics in QCD\,\cite{JQVY06,JQVY06DIS}.

In the present work, we shall focus on the twist-3 mechanism. 
A systematic study on this mechanism was first performed in \cite{QS91}
for the direct photon production $p^\uparrow p\to \gamma X$,
and the method has been applied to many other processes such as
pion production in $pp$-collision, 
$p^\uparrow p\to\pi X$\,\cite{QS99,KK00,Koike03,KQVY06}, 
hyperon polarization $pp\to\Lambda^\uparrow X$\,\cite{KK01,hyperon},
Drell-Yan lepton-pair production $p^\uparrow p\to \ell^+ \ell^- X$\,\cite{JQVY06}, 
pion production in semi-inclusive deep inelastic scattering
(SIDIS), $ep^\uparrow\to e\pi X$\,\cite{ekt06,JQVY06DIS,EKT}.
In our recent work\,\cite{ekt06}, we reexamined the formalism for the twist-3 mechanism
and gave a proof 
for the factorization property and the gauge invariance 
of the corresponding single-spin-dependent cross sections, which was missing in 
the previous literature.  Through this development, 
the cross section formula derived in the previous studies\,\cite{QS91}-\cite{EKT}
have been given a solid theoretical basis. 

In the twist-3 mechanism, 
the strong interaction phase necessary for SSA 
is provided by the partonic hard scattering:
Owing to the insertion
of a ``coherent gluon'' emanating from the twist-3 
quark-gluon correlation 
inside e.g. the polarized nucleon, 
an internal propagator of the partonic hard part
can be on-shell, and its imaginary part (pole contribution)
can give rise to the interfering phase that leads eventually to 
the real single-spin-dependent cross section. 
Depending on the resulting value of the coherent-gluon's
momentum fraction at such poles,
those poles are classified as soft-gluon-pole (SGP), soft-fermion-pole (SFP) and
the hard-pole (HP).  
In \cite{ekt06}, we have given a formula which expresses 
all these pole contributions in terms of the
twist-3 distributions associated with the gluon's field strength tensor 
(see (\ref{twist3distr}), (\ref{w-twist3-f}) below). 
There we have also given a general proof that the 
SGP contribution appears as both ``derivative'' and ``non-derivative'' 
terms of the twist-3 correlation functions,
while the SFP and HP contributions appear only as the non-derivative terms. 
Since the derivative of the ``SGP function'', a twist-3 correlation function at the SGP,
causes enhancement compared with the non-derivative term in a certain kinematic region,
and also one may expect that the soft gluons are more ample in the hadrons, 
phenomenology based on the SGP contribution only has often been performed in the 
literature\,\cite{QS99}-\cite{EKT}.

Another peculiar feature of the SGP contribution is that
the partonic hard cross section associated with the ``derivative term'', 
arising from the twist-3 soft-gluon mechanism,
is directly proportional to some twist-2 partonic cross section 
as noticed for direct photon production\,\cite{QS91}, 
Drell-Yan pair production\,\cite{JQVY06}, 
SIDIS\,\cite{EKT}, $p^\uparrow p\to\pi X$\,\cite{QS99,KK00,Koike03} 
and $pp\to\Lambda^\uparrow X$\,\cite{KK01,hyperon}. 

In this Letter we propose a new systematic approach to treat the SGPs,
which allows us to reveal the novel structure behind the soft-gluon mechanism to the SSA.
We disentangle the coupling of the 
coherent gluon and the associated pole structure 
from the partonic subprocess, by reorganizing 
the relevant diagrams
using Ward identities and certain decomposition identities
for the interacting 
parton propagators. 
We find that the many Feynman diagrams can be eventually united 
into certain derivative of the Born diagrams without the coherent-gluon insertion,
which shows that the entire contributions from the SGPs, not only the derivative term but also
the non-derivative term,  
can be derived 
from the knowledge of the twist-2 cross section formula for the unpolarized process.
We establish the corresponding ``master formula'' that 
is applicable to 
a range of processes 
like SIDIS, Drell-Yan and direct-photon production.

To illustrate our approach, we consider the SIDIS, 
$e(\ell)+p(p, S_{\perp})\to e(\ell')+\pi(P_h)+X$,
following the convention 
of our recent work\,\cite{ekt06}: We use
the kinematic variables, $S_{ep}=(p + \ell)^2$,
$q=\ell-\ell'$, $Q^2 =-q^2$, $x_{bj}={Q^2/ (2p\cdot q)}$, and
$z_f={p\cdot P_h / p\cdot q }$. 
All momenta $p, P_h, \ell$, and $\ell'$ of the particles in the initial and the final states
can be regarded as light-like
in the twist-3 accuracy, $p^2=P_h^2 =\ell^2 ={\ell'}^2 =0$.
As usual, we define another light-like vector as
$n^\mu = (q^\mu + x_{bj} p^\mu )/p\cdot q$, and the projector onto the transverse 
direction as $g_{\perp}^{\mu \nu} = g^{\mu \nu} - p^\mu n^\nu - p^\nu n^\mu$.
We also define a space-like vector, 
$q_t^\mu=q^\mu- ({P_h\cdot q/ p\cdot P_h}) p^\mu -
({p\cdot q/ p\cdot P_h})P_h^\mu$,
which is orthogonal to
both $p$ and $P_h$, 
and its magnitude as
$q_T = \sqrt{-q_t^2}$.
Then, in a frame where the
3-momenta $\vec{q}$ and $\vec{p}$ of the virtual photon 
and the initial nucleon are collinear along the 
$z$-axis, 
like the so-called hadron frame~\cite{MOS92},
the magnitude of the transverse momentum of the pion
is given by $\sqrt{-P_{h\perp}^2}=z_f q_T$. 

In the present study we are interested in the SSA for the large-$P_{h\perp}$ pion production,
in particular, the contribution from the twist-3 quark-gluon correlation functions for
the transversely polarized nucleon,
which are defined as\,\cite{QS91,KT99}
\begin{eqnarray}
M^{\beta}_{Fij} (x_1,x_2)
& =&\int {d\lambda\over 2\pi}\int{d\mu\over 2\pi}
e^{i\lambda x_1}e^{i\mu(x_2-x_1)}
\langle p\ S_\perp |\bar{\psi}_j(0)[0,\mu n]
{gF^{\beta\rho}(\mu n)n_\rho}[\mu n, \lambda n]
\psi_i(\lambda n)|p\ S_\perp \rangle\nonumber\\
& =&
{M_N\over 4} \left(\pslash\right)_{ij} 
\epsilon^{\beta pnS_\perp}
{G_F(x_1,x_2)}+ 
i{M_N\over 4} \left(\gamma_5\pslash\right)_{ij} 
S_\perp^\beta
{\GFt(x_1,x_2)}+ \cdots,
\label{twist3distr}
\end{eqnarray}
where 
the spinor indices $i$ and $j$ associated with the quark field $\psi$ 
are shown explicitly,
$F^{\beta \rho}$ is the gluon field strength tensor,
and the spin vector for the transversely polarized nucleon
satisfies $S_\perp^\alpha = g_{\perp}^{\alpha \beta} S_{\perp \beta}$, $S_\perp^2 = -1$.
The path-ordered gauge-link, 
$[\mu n,\lambda n]={\rm P} \exp\left[ ig \int_{\lambda}^{\mu}dt n\cdot A(tn) \right]$,
guarantees gauge invariance of the nonlocal lightcone operator.
The second line of (\ref{twist3distr}) defines two dimensionless functions
$G_F(x_1,x_2)$ and $\GFt(x_1,x_2)$ through the Lorentz decomposition of the matrix element;
here $M_N$ is the nucleon mass representing typical mass scale generated by nonperturbative effects, 
and ``$\cdots$'' denotes Lorentz structures of twist higher than three.
These two functions $G_F(x_1,x_2)$ and $\GFt(x_1,x_2)$ constitute a
complete set of the twist-3 quark-gluon correlation functions for the transversely polarized 
nucleon\,\cite{EKT,KT99}.

\begin{figure}[t!]
\begin{center}
\epsfig{figure=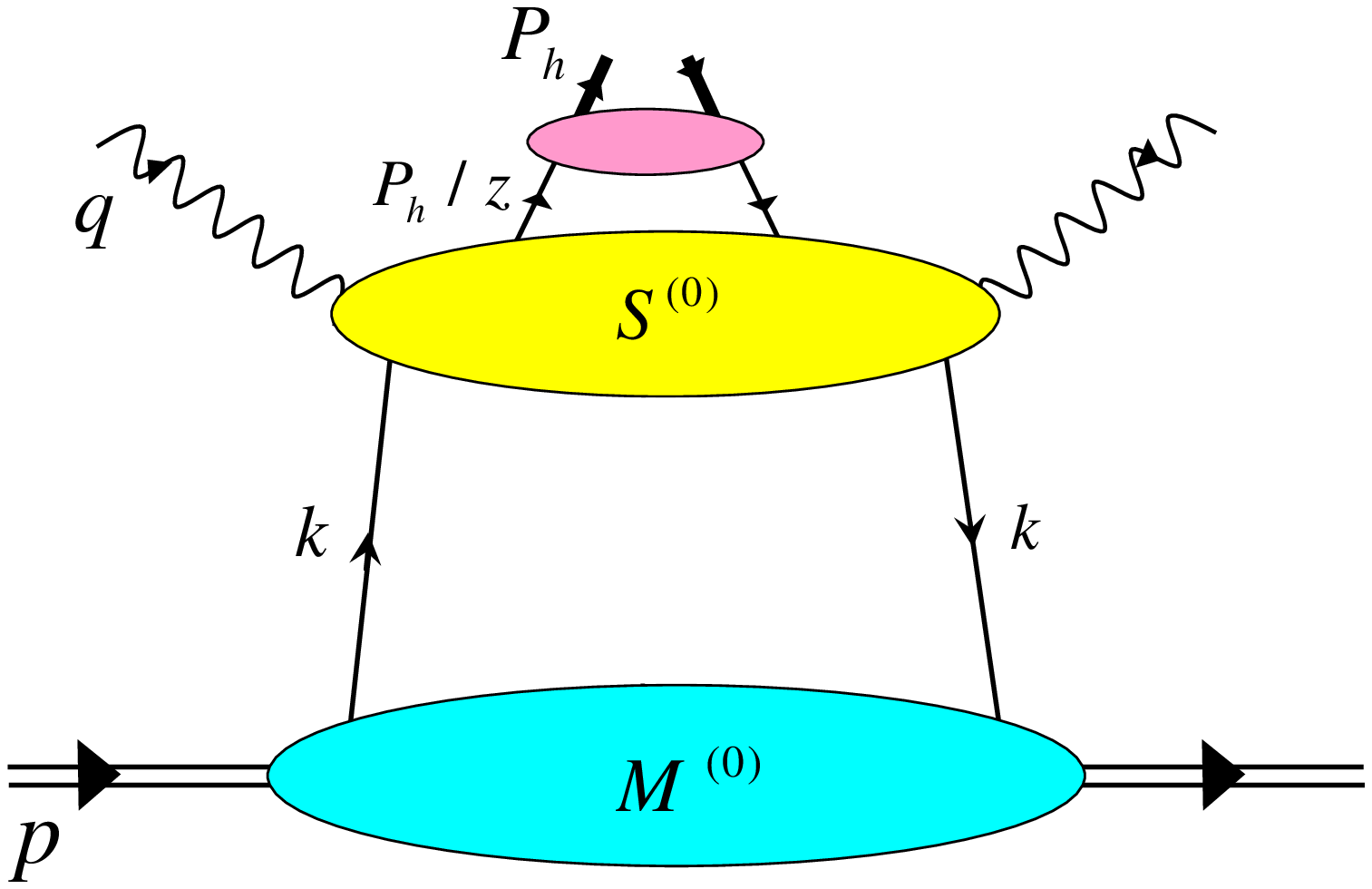,width=0.4\textwidth,clip=}
\hspace{22mm}
\epsfig{figure=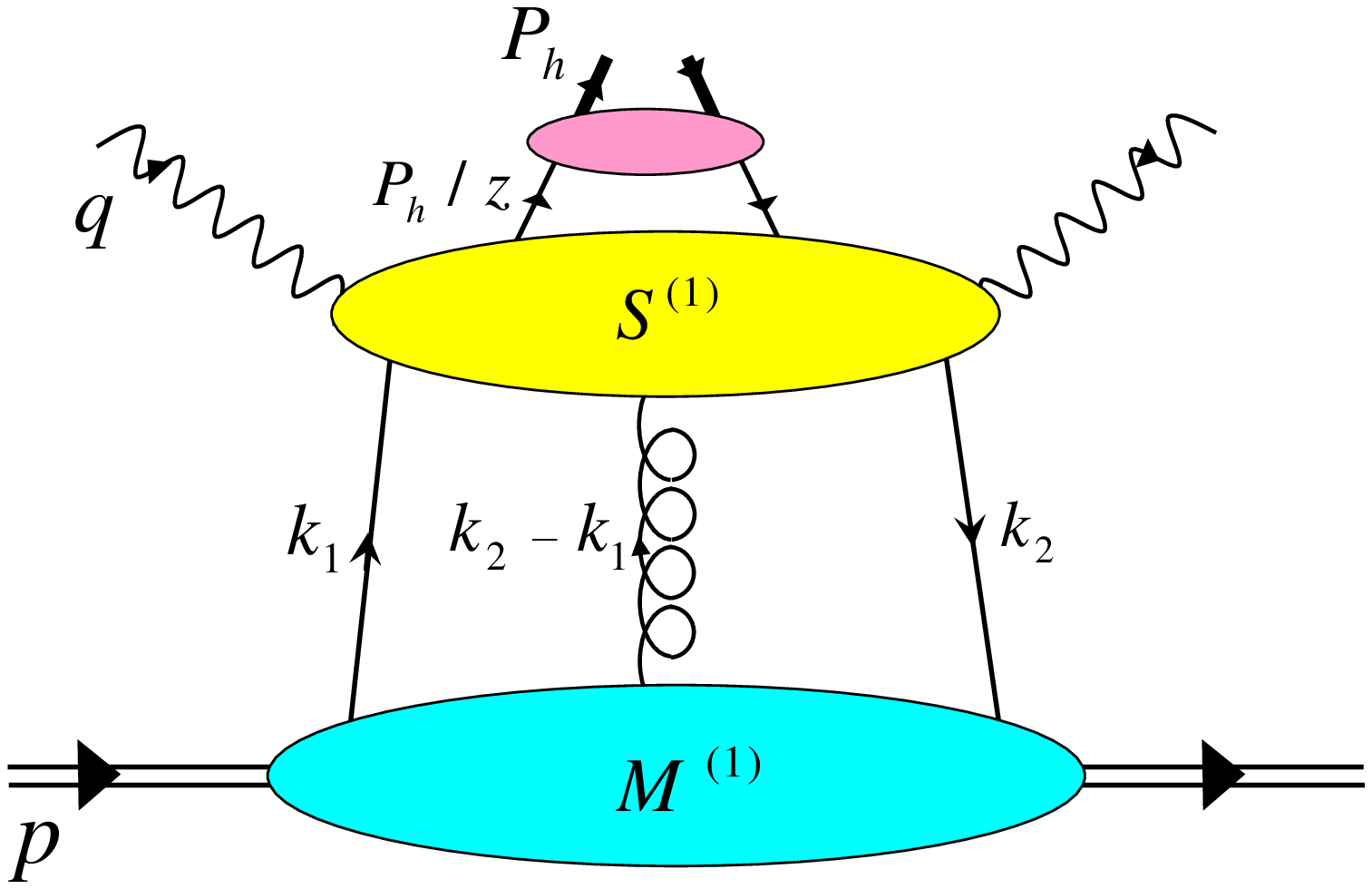,width=0.4\textwidth,clip=}
\end{center}
\vspace{-0.5cm}
\hspace{3.4cm}(a)
\hspace{8.4cm}(b)
\caption{Generic diagrams for the hadronic tensor of $ep^\uparrow\to e\pi X$,
decomposed into the three blobs as nucleon matrix element (lower), pion matrix element (upper),
and partonic hard scattering by the virtual photon (middle). The first two terms, (a) and (b),  
in the expansion by the number of partons connecting the middle and lower blobs
are relevant to the twist-3 effect induced by the nucleon.
\label{fig:s1}}
\vspace*{0.cm}
\end{figure}

The relevant contributions to the hadronic tensor $W_{\mu \nu}$ arise from the process 
where the partons from the nucleon in the initial state
undergoes the hard interaction with the virtual photon,
followed by the fragmentation into the final state with $\pi + {\rm anything}$,
as illustrated in Fig.~\ref{fig:s1}.
The twist-3 distribution functions contribute to $ep^\uparrow\to e\pi X$ in combination with
the twist-2 
fragmentation function for the pion, which is immediately factorized from the 
hadronic tensor as
\beq
W_{\mu\nu}(p,q,P_h)=\sum_{j=q,g}\int{dz\over z^2}D_j(z) w^j_{\mu\nu}(p,q,{P_h\over z})\ ,
\label{Wmunu}
\eeq
where $D_j(z)$ ($j=q,\ g$) is the quark and gluon fragmentation functions for the pion, 
with $z$ being the momentum fraction.  
We consider the case for the quark fragmentation in detail and
omit the index $j$ from $w_{\mu\nu}^j$ below.  Modifications necessary for the gluon-fragmentation
case will be discussed later.
The lower blobs in Fig. 1 can be written as Fourier transform of the
correlation functions
for the nucleon, i.e., schematically, $M^{(0)}(k)\sim \la pS_\perp|\bar{\psi}\psi|pS_\perp\ra$
and 
$M^{(1)\sigma}(k_1,k_2)\sim \la pS_\perp|\bar{\psi}A^\sigma\psi|pS_\perp\ra$,
with the upper indices (0) and (1) representing the number of gluon lines connecting
the middle and lower blobs.  
(See Eqs.~(23), (24) of \cite{ekt06} for the explicit definitions.)
The momenta of the partons, $k$ and $k_{1,2}$, are assigned as in Fig.~1.

In the leading-order perturbation theory for the partonic hard scattering,
we have shown\,\cite{ekt06} that 
the twist-3 contribution to $w_{\mu\nu}$ 
arises 
solely from Fig.~1~(b), and the entire twist-3 contribution can be written 
as
\begin{equation}
w_{\mu\nu}(p,q,\frac{P_h}{z}) 
=\int dx_1\int dx_2{\rm Tr}\left[ i
g_{\perp \beta}^{\alpha}
M_{F}^{\beta} (x_1,x_2)
\left. {\partial S^{(1)}(k_1,k_2, P_h /z) 
\over \partial k_{2\perp}^\alpha}  \right|_{k_i=x_ip}
\right] ,
\label{w-twist3-f}
\end{equation}
where 
$M_{F}^{\beta} (x_1,x_2)$ is defined in 
(\ref{twist3distr}), and 
$S^{(1)}(k_1,k_2, P_h /z) \equiv S^{(1)}_\sigma (k_1,k_2,{P_h / z}) p^\sigma$
represents the middle blob in Fig.~\ref{fig:s1}~(b), 
which denotes the hard scattering
function for the partons off the virtual photon, corresponding to 
the nucleon matrix element $M^{(1)\sigma}(k_1,k_2)$.  
For simplicity, we suppress 
momentum $q$ and the Lorentz indices 
$\mu$, $\nu$ for the virtual photon in $S^{(1)}_\sigma(k_1,k_2, P_h /z)$.
In (\ref{w-twist3-f}), $S^{(1)}(k_1,k_2, {P_h / z})$ and $M_{F}^{\beta} (x_1,x_2)$
are matrices in spinor space, and
${\rm Tr}[\cdots ]$ indicates the trace over 
Dirac-spinor indices while the color trace is implicit. 
Here we recall from \cite{ekt06} that, 
although the straightforward collinear expansion to the twist-3 
accuracy produces many other complicated terms, it has been proved 
that the hard parts associated with those terms 
vanish in the leading order in QCD perturbation theory, 
using Ward identities for color gauge invariance. 
Note that (\ref{w-twist3-f}) guarantees the factorization property for the twist-3 
single-spin-dependent cross section in manifestly gauge-invariant form,
which was assumed in the previous literature\,\cite{JQVY06DIS,EKT} (see
also \cite{QS91,JQVY06}).
We should also mention 
that, for the HP and SFP contributions, one can calculate the partonic hard scattering function
via $S^{(1)}_\alpha(x_1p,x_2p, P_h /z)$
by using the relation 
$\left. {\partial S^{(1)}(k_1,k_2, P_h /z) / \partial k_{2\perp}^\alpha}  \right|_{k_i=x_ip}
=S^{(1)}_\alpha(x_1p,x_2p, P_h /z)/(x_1-x_2)$
obtained from the Ward identity, while, for the SGP contribution,
one has to calculate the ``derivative'',
$\left. {\partial S^{(1)}(k_1,k_2, P_h /z) / \partial k_{2\perp}^\alpha}  \right|_{k_i=x_ip}$, 
as shown in (\ref{w-twist3-f})\,\cite{ekt06}.

\begin{figure}[t!]
\begin{center}
\epsfig{figure=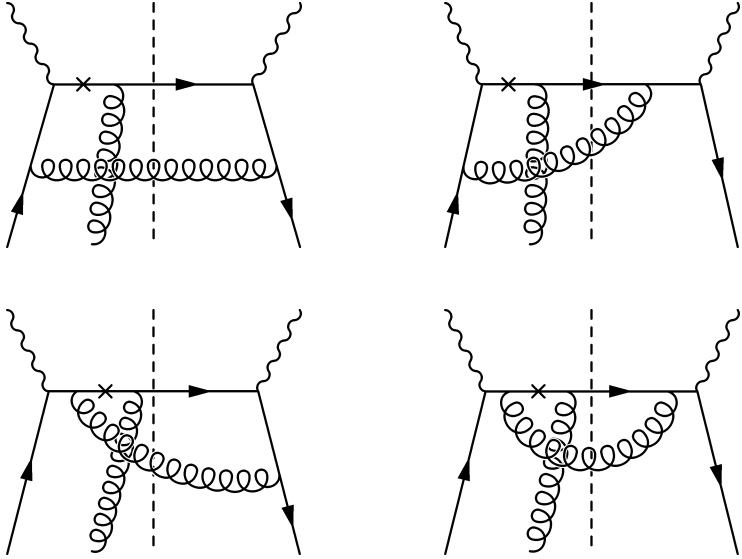,width=0.6\textwidth}
\end{center}
\caption{Feynman diagrams which give rise to the 
SGP contributions in the quark fragmentation channel, where
the hard quark
fragments into the final-state with pion and the hard gluon goes into unobserved final state. 
The cross $\times$ denotes 
the quark propagator which gives the SGP contribution.
Mirror diagrams also contribute.
\label{fig1}
}
\end{figure}

\begin{figure}[t!]
\begin{center}
\epsfig{figure=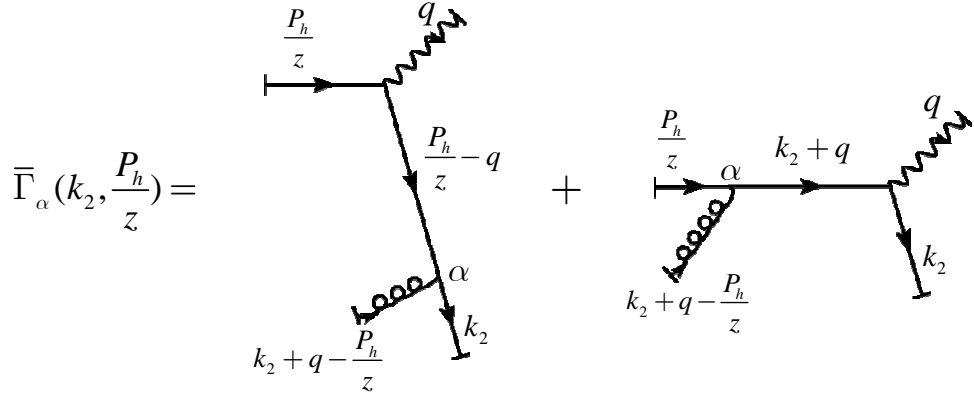,width=0.9\textwidth,clip=}
\end{center}
\caption{The definition of the photon-quark-gluon
vertex function $\bar{\Gamma}_\alpha  (k_2 , P_h /z)$, 
as the sum of two types of diagrams 
which appear in the RHS of the cut in Fig.~\ref{fig1}.
The amputated lines are identified by bars at their ends.
\label{fig3}
}
\end{figure}

Based on (\ref{w-twist3-f}), our task is to identify the SGP contribution to
$S^{(1)}(k_1,k_2,{P_h / z})$ and compute its deviation arising 
linearly in the quark's transverse momentum $k_{2\perp}$ from the value in the
collinear limit $k_{1,2} \rightarrow x_{1,2} p$.
For this purpose, we work in the Feynman gauge and with 
$k_{i \perp} \ll x_i p$ 
in $k_i = x_i p + (k_i \cdot p) n + k_{i \perp}$ ($i=1,2$).
In the leading order in QCD perturbation theory, 
$S^{(1)}(k_1,k_2, P_h / z)$ 
stands for a set of cut Feynman diagrams which are obtained by
attaching the additional 
gluon to the 2-to-2 partonic Born subprocess,
where the large transverse momentum $P_{h\perp}/z$ of the ``fragmenting quark'' 
is provided
by the recoil from the emission of a hard gluon into the final state. 
When the coherent gluon couples to an 
on-shell parton line, 
the parton propagator adjacent to the coherent gluon produces a pole for the vanishing gluon momentum,
$k_2 -k_1 \rightarrow 0$.
Only those arising from the diagrams in Fig.~\ref{fig1}, 
where the coherent gluon couples to the final-state quark line
fragmenting into $\pi + {\rm anything}$, survive as the SGP contributions\,\cite{EKT,ekt06,JQVY06DIS}, 
while the other pole contributions cancel out combined with those
from the ``mirror'' diagrams. 
We denote the contributions of the diagrams in Fig.~\ref{fig1}, 
where the coherent gluon is attached to the LHS of the cut,
as $S^{(1)L}(k_1,k_2,{P_h / z})$, and those of the mirror diagrams
as $S^{(1)R}(k_1,k_2,{P_h / z})$, so that
$S^{(1)}(k_1,k_2,{P_h / z}) =S^{(1)L}(k_1,k_2,{P_h / z}) +
S^{(1)R}(k_1,k_2,{P_h / z})$. 
Explicit form of $S^{(1)L}(k_1,k_2,{P_h / z})$ is given as
\begin{eqnarray}
S^{(1)L}(k_1,k_2 , \frac{P_h}{z} )  &=& \frac{-1}{2N_c} \bar{\Gamma}_{\alpha}(k_2 , \frac{P_h}{z}  )
\frac{\slash{P}_h}{z}i\gamma_\sigma p^\sigma 
\frac{i}{\frac{\slash{P}_h}{z} +\slash{k}_1 - \slash{k}_2 + i\varepsilon}
\Gamma_\beta (k_1, \frac{P_h}{z} + k_1 - k_2 )
\nonumber\\
&\times&
{\cal D}_{+}^{\alpha \beta}(k_2 + q -\frac{P_h}{z})\ ,
\label{sL}
\end{eqnarray}
where 
$\bar{\Gamma}_{\alpha} (k_2 , P_h /z)\equiv \gamma^0 \Gamma_\alpha^\dagger (k_2 ,P_h /z)\gamma^0$ 
denotes the photon-quark-gluon
vertex function of Fig.~\ref{fig3} which appears in the RHS of the cut in the diagrams 
of Fig.~\ref{fig1}. 
In $\bar{\Gamma}_{\alpha}(k_2,P_h/z)$, the factors for the external lines are amputated. 
With this $\Gamma_\alpha$, 
the photon-quark-gluon
vertex function appearing 
in the LHS of the cut in Fig.~\ref{fig1} is given by 
$\Gamma_\beta (k_1, P_h /z + k_1 - k_2 )$. 
The structure $\slash{P}_h /z$ projects the final-state quark onto the twist-2 fragmentation 
process to produce the pion (see (\ref{Wmunu})),
and $i\gamma_\sigma p^\sigma$ is the quark-coherent-gluon vertex.
${\cal D}_{+}^{\alpha \beta}(k)= 2\pi
\delta\left(k^2 \right) \theta (k^0 )\sum_\lambda
\epsilon_{(\lambda)}^\alpha(k)\epsilon_{(\lambda)}^{*\beta}(k)$
is the cut propagator for the final-state hard gluon with the polarization vector
$\epsilon_{(\lambda)}^\alpha(k)$.
Note that we have already worked out the color structure associated with (\ref{w-twist3-f}), 
simplifying the color matrices contained in Fig.~\ref{fig1} as $t^a t^b t^a = (-1/2N_c )t^b$ 
and performing the color trace to obtain the gauge-invariant matrix element (\ref{twist3distr}).
We also note the relation 
$S^{(1)R} (k_1,k_2,{P_h / z})  = \bar{S}^{(1)L} (k_2,k_1,{P_h / z})
\equiv  \gamma^0 {S^{(1)L}}^{\dagger} (k_2 , k_1 , P_h /z) \gamma^0$. 

To proceed further, one may employ direct evaluation of each diagram of Fig.~\ref{fig1}, 
substituting the explicit form corresponding to Fig.~\ref{fig3} 
into the vertex functions
$\bar{\Gamma}_{\alpha}(k_2 , P_h /z)$ and $\Gamma_\beta (k_1, P_h /z + k_1 - k_2 )$
of (\ref{sL}), and  working out the
necessary Dirac algebra.
Such calculation has been done in \,\cite{ekt06,JQVY06DIS,EKT}.
Here we employ another approach to disentangle the coherent-gluon vertex 
and the corresponding SGP
structure from the diagrams of Fig.~\ref{fig1}.
An earlier attempt 
in the same spirit investigated decoupling property
of the soft-gluon vertex in the collinear limit, $k_{1\perp}=k_{2\perp}=0$,
for the SGP contribution in hadron-hadron scattering,
using the helicity basis technique\,\cite{Ratcliffe:1998pq};
but fully consistent evaluation of the SGP contribution following \cite{ekt06}
requires to 
treat the deviation of the hard part
due to nonzero $k_{2\perp}$, as noted above, and in this case 
straightforward application of the helicity basis technique would be useless.
Our new systematic approach uses Ward identities and decomposition
identities for the  
parton propagators interacting with the coherent gluon,
which allows us to disentangle
the coherent gluon vertex from the derivative 
$\left. 
{\partial S^{(1)}(k_1,k_2, P_h /z) 
/ \partial k_{2\perp}^\alpha}  \right|_{k_i=x_ip}$ in (\ref{w-twist3-f}).
A key idea is to reorganize the terms that contribute to this derivative
by rewriting $p^\sigma$ contracted 
with the quark-gluon vertex in (\ref{sL}) as
\begin{equation}
p^\sigma = \frac{1}{x_2 - x_1 - i\varepsilon} 
\left(k_2^\sigma - k_1^\sigma \right)  - 
\frac{1}{x_2 - x_1 - i\varepsilon} 
\left(k_{2\perp}^\sigma - k_{1\perp}^\sigma \right)  \ ,
\label{psigma}
\end{equation}
up to the irrelevant $O( \left(k_{2\perp} - k_{1\perp}\right)^2 )$ correction.
Correspondingly,
$S^{(1)L}(k_1,k_2,P_h/z)$ can be decomposed as
\begin{equation}
S^{(1)L}(k_1,k_2,\frac{P_h}{z}) = S^{(1\cdot)L}(k_1,k_2 ,\frac{P_h}{z} ) 
+S^{(1\perp)L}(k_1,k_2,\frac{P_h}{z})\ ,
\label{deco}
\end{equation}
where the first and second terms in the RHS 
correspond
to those in (\ref{psigma}), respectively.
In (\ref{psigma}), ``$-i\varepsilon$'' in the denominator is chosen such that
each term in (\ref{deco}) does not produce pinch singularity
at $x_1=x_2$. 

\begin{figure}[t!]
\begin{center}
\epsfig{figure=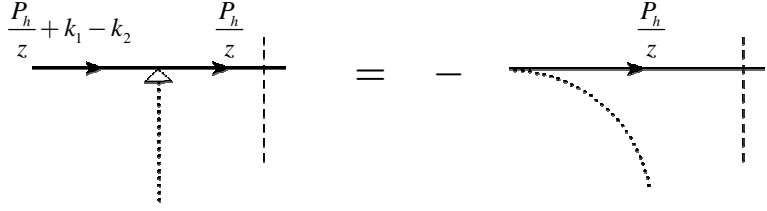,width=0.8\textwidth,clip=}
\end{center}
\caption{Diagrammatic representation of Ward identity for the coupling of the 
scalar-polarized gluon to the final-state quark line.
\label{fig:Ward}
}
\end{figure}

We can exploit an elementary Ward identity in $S^{(1\cdot)L}(k_1,k_2,{P_h / z})$
in order to disentangle the scalar-polarized gluon vertex, $\slash{k}_2 - \slash{k}_1$, 
as well as the quark propagator adjacent to it, as
$(\slash{P}_h /z)(\slash{k}_2 - \slash{k}_1 )
[1/ (\slash{P}_h /z +\slash{k}_1 - \slash{k}_2 + i\varepsilon)] 
= (\slash{P}_h /z)(\slash{k}_2 - \slash{k}_1 - \slash{P}_h /z )
[1/ (\slash{P}_h /z +\slash{k}_1 - \slash{k}_2 + i\varepsilon)] 
=-\slash{P}_h /z$, 
as illustrated in Fig.~\ref{fig:Ward},
and get
\begin{equation}
S^{(1\cdot)L}(k_1,k_2 ,\frac{P_h}{z} ) = 
\frac{-1}{2N_c}
\frac{1}{x_2 - x_1 - i\varepsilon}
 \bar{\Gamma}_{\alpha}(k_2 ,\frac{P_h}{z} )
\frac{\slash{P}_h}{z}
\Gamma_\beta (k_1, \frac{P_h}{z} + k_1 - k_2 )
{\cal D}_{+}^{\alpha \beta}(k_2 + q -\frac{P_h}{z})\ .
\label{Phi}
\end{equation}
This exhibits the SGP only as the single pole, so that we can put $x_2 = x_1$
except for the factor $1/( x_2 - x_1 - i\varepsilon)$,
without affecting the results for the SGP contribution.
Combining this with the corresponding result for 
$S^{(1\cdot)R}(k_1 , k_2 , {P_h / z}) =
\bar{S}^{(1\cdot)L}(k_2 , k_1 , {P_h / z})$
and Taylor expanding 
the total result
$S^{(1\cdot)}(k_1,k_2,{P_h / z})$$=S^{(1\cdot)L}(k_1,k_2,{P_h / z})$ 
$+ S^{(1\cdot)R}(k_1 , k_2 , {P_h / z})$ in terms of $k_{1\perp}$ and $k_{2 \perp}$, we get
\begin{equation}
S^{(1\cdot)}(k_1,k_2, \frac{P_h}{z})=
\frac{1}{2N_c}
\frac{k_{2\perp}^\sigma -k_{1\perp}^\sigma}{x_2 - x_1 - i\varepsilon}
 \left. \frac{\partial}{\partial r^{\sigma}_\perp} \bar{\Gamma}_{\alpha}(x_1 p , r)
\frac{\slash{P}_h}{z}
\Gamma_{\beta} (x_1 p , r)
{\cal D}_{+}^{\alpha \beta}(x_1 p + q -r )
\right|_{r\rightarrow \frac{P_h}{z}} ,
\label{ward2}
\end{equation}
up to the irrelevant terms of the second or higher
order in $k_{1\perp}, k_{2\perp}$. Here
$r$ denotes a four-vector not restricted to being light-like, $r^2 \neq 0$.

Next we consider the contribution from $S^{(1\perp)L}(k_1,k_2,{P_h / z})$ 
of the second term of (\ref{deco}).
Because the coherent gluon vertex associated with $S^{(1\perp)L}(k_1,k_2,{P_h / z})$ is proportional 
to $k_{2\perp}^\sigma -k_{1\perp}^\sigma$ (see (\ref{sL}), (\ref{psigma})), we can set
$k_{1\perp}=k_{2\perp}=0$ in $S^{(1\perp)L}(k_1,k_2,{P_h / z})$ except for this vertex factor,
up to irrelevant corrections.  
One thus obtains
\begin{eqnarray}
S^{(1\perp)L}(k_1,k_2 , \frac{P_h}{z} )\!\! & &= \frac{-1}{2N_c}
\frac{k_{2\perp}^\sigma -k_{1\perp}^\sigma}{x_2 - x_1 -i\varepsilon} 
\bar{\Gamma}_{\alpha}(x_2 p , \frac{P_h}{z}  )
\nonumber\\
\times 
\frac{\slash{P}_h}{z}\gamma_\sigma & &\!\!\!\!
\frac{1}{\frac{\slash{P}_h}{z} + (x_1 - x_2 )\slash{p} + i\varepsilon}
\Gamma_\beta (x_1 p, \frac{P_h}{z} + x_1 p - x_2  p )
{\cal D}_{+}^{\alpha \beta}(x_2 p + q -\frac{P_h}{z})\ .
\label{sL2}
\end{eqnarray}
In this result we decompose the product of the quark-gluon vertex and the quark propagator, 
adjacent to
the fragmentation insertion $\slash{P}_h /z$, as
\begin{equation}
\frac{\slash{P}_h}{z} \gamma_{\sigma}
\frac{\slash{P}_h /z +( x_1 -x_2 )\slash{p}}{\left(\frac{P_h}{z} +(x_1  -x_2 ) p\right)^2 + i\varepsilon} 
= \frac{\slash{P}_h}{z} \left( -\frac{P_{h \sigma}}{P_h \cdot p } \frac{1}{x_2 - x_1  - i\varepsilon}
+ \gamma_\sigma \frac{z\slash{p}}{2P_h \cdot p } \right) ,
\label{eiko}
\end{equation} 
where the first term in the parenthesis is given by the eikonal vertex and
eikonal propagator, and the second term is the ``contact term''. 
Combining the result obtained from (\ref{sL2}), (\ref{eiko}) 
with the corresponding result for
$S^{(1\perp)R}(k_1 , k_2 , {P_h / z}) =
\bar{S}^{(1\perp)L}(k_2 , k_1 , {P_h / z})$,
we find that 
the double pole term
in $S^{(1\perp)L}(k_1,k_2,{P_h / z})$, 
which originates from the eikonal propagator in (\ref{eiko})
and gives the contribution proportional to $\delta'(x_1-x_2)$,
cancels the corresponding double-pole 
term in $S^{(1\perp)R}(k_1 , k_2 , {P_h / z})$, and the remaining single-pole contributions 
eventually give, for 
$S^{(1\perp)}(k_1,k_2,{P_h / z})  =S^{(1\perp)L}(k_1,k_2,{P_h / z}) 
+ S^{(1\perp)R}(k_1 , k_2 , {P_h / z})$, 
\begin{eqnarray}
\lefteqn{
S^{(1\perp)}(k_1,k_2, P_h /z)
=
\frac{1}{2N_c}\frac{k_{2\perp}^\sigma -k_{1\perp}^\sigma}{x_2 - x_1 - i\varepsilon} 
\left. \frac{\partial}{\partial r_{\rho}} \right\{ 
g_{\sigma \rho} \bar{\Gamma}_{\alpha}(x_1 p , \frac{P_h}{z} )
\slash{r}
\Gamma_{\beta} (x_1 p , \frac{P_h}{z} )}
\nonumber\\
&&
\;\;\;\;\;\;\;\;
\times
{\cal D}_{+}^{\alpha \beta}(x_1 p + q -\frac{P_h}{z}  )
- 
\left.\left.
\frac{P_{h\sigma}p_{\rho}}{P_{h}\cdot p}
\bar{\Gamma}_{\alpha}(x_1 p , r)
\slash{r}
\Gamma_{\beta} (x_1 p , r)
{\cal D}_{+}^{\alpha \beta}(x_1 p + q -r )\right\}
\right|_{r\rightarrow \frac{P_h}{z}} .
\label{perppart}
\end{eqnarray}

Combining (\ref{ward2}) and (\ref{perppart}), the entire SGP contribution 
for (\ref{w-twist3-f}) reads~\footnote{
We also obtain $\left. 
(\partial /\partial k_{1\perp}^{\rho} + \partial /\partial k_{2\perp}^{\rho} ) 
S^{(1)}(k_1,k_2, P_h /z) \right|_{k_i=x_ip} = 0$ for the SGP contribution, 
which has been proved in \cite{ekt06}
by the detailed inspection of the diagrams in Fig.~\ref{fig1}.
This holds for the gluon fragmentation channel, too (see (\ref{wardg2})).}
\begin{equation}
\left. \frac{\partial S^{(1)}(k_1,k_2, P_h /z)}{\partial k_{2\perp}^{\sigma}}  
\right|_{k_i=x_ip} =
\frac{1}{2N_c C_F}
\frac{z}{x_2 - x_1 - i\varepsilon}
\frac{\partial S^{(0)}(x_1 p, \frac{P_h}{z})}{\partial P_{h \perp}^\sigma}\ ,
\label{total}
\end{equation}
where $C_F = (N_c^2 -1)/(2N_c)$ and 
\begin{equation}
S^{(0)}(x_1 p, \frac{P_h}{z} )=C_F \bar{\Gamma}_{\alpha}(x_1 p , \frac{P_h}{z} )
\frac{\slash{P}_h}{z}
\Gamma_{\beta} (x_1 p , \frac{P_h}{z} )
{\cal D}_{+}^{\alpha \beta}(x_1 p + q -\frac{P_h}{z}  )
\label{s0}
\end{equation}
is exactly the leading-order hard scattering function in the collinear limit $k \rightarrow xp$,
which is represented as the middle blob in Fig.~1~(a).
In (\ref{total}), 
we have used the relation,
\begin{equation}
\left( g_{\sigma \rho}-\frac{P_{h \sigma}p_{\rho}}{P_{h}\cdot p} \right)
\left.  \frac{\partial \varphi(r)}{\partial r_{\rho}} \right|_{r\rightarrow \frac{P_h}{z}}
= z \frac{\partial \varphi(P_h /z )}{\partial P_{h}^\sigma}\ ,
\label{eq:relrel}
\end{equation}
for $P_h = -(P_{h\perp}^2/2P_h \cdot p)p+(P_h \cdot p)n+P_{h\perp}$,
which holds for 
an arbitrary function $\varphi(r)$.  
Note that $P_h\cdot n$ is not an independent variable 
to perform the derivative in the RHS of (\ref{eq:relrel}),
corresponding to that the LHS vanishes when contracted by $p^\sigma$.
Substituting (\ref{twist3distr}) and (\ref{total})
into (\ref{w-twist3-f}), and integrating 
over $x_2$ to get the imaginary phase from the SGP of (\ref{total}),
one gets for $w_{\mu\nu}$ as 
\begin{equation}
w_{\mu \nu} = \frac{1}{2N_c C_F}\left(\frac{-\pi M_N z}{4}\right) 
\epsilon^{\sigma pnS_\perp}\frac{\partial}{\partial P_{h \perp}^\sigma} 
\int dx
G_F(x,x) 
{\rm Tr}\left[ S^{(0)} (x p, \frac{P_h}{z})\pslash \right]  + \cdots\ ,
\label{wmunu}
\end{equation}
where 
the ellipses stand for the contribution from 
the second term in the RHS of (\ref{twist3distr}),
which does not contribute to the cross section when contracted with
the symmetric leptonic tensor for the unpolarized electron.
We recall the similar formula
for the leading-order twist-2 hadronic tensor,
\begin{equation}
w_{\mu \nu}^{\mbox{\scriptsize tw-2}}= \frac{1}{2}
\int dx
f_q(x) 
{\rm Tr}\left[ S^{(0)} (x p, \frac{P_h}{z})\pslash \right] + \cdots\ ,
\label{wmunutw2}
\end{equation}
where 
$S^{(0)}(xp,P_h/z)$ in (\ref{wmunu}) 
appears and
$f_q(x)$ is the unpolarized quark distribution.  

\begin{figure}[t!]
\begin{center}
\epsfig{figure=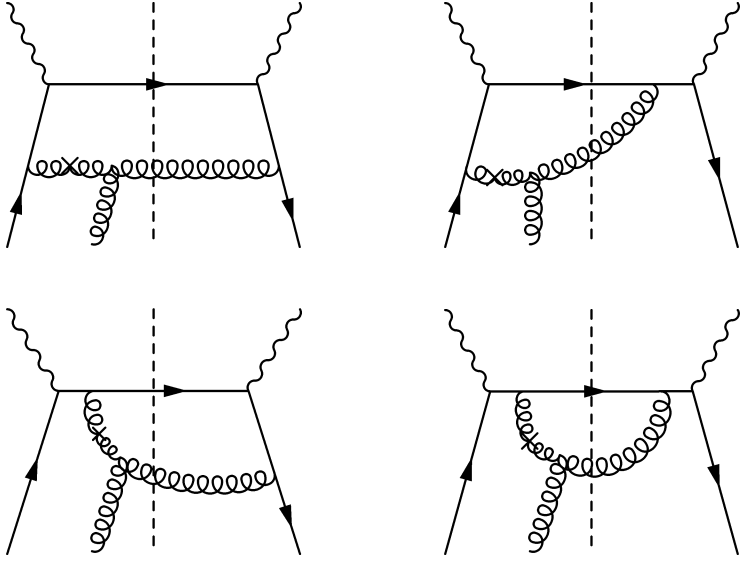,width=0.6\textwidth}
\end{center}
\caption{Same as Fig.~\ref{fig1}, but for the gluon fragmentation channel
where the hard gluon 
fragments into the final-state with pion and the hard quark goes into unobserved final state. 
\label{fig:gluon}
}
\end{figure}

It is straightforward to extend the above results to the gluon fragmentation case
by mostly trivial substitutions,
noting that essential part of our above derivation is based on diagrammatic manipulation.
The diagrams for the SGP contribution in this case are shown 
in Fig.~\ref{fig:gluon}\,\cite{ekt06,EKT}. 
The corresponding hard scattering function is given by
\begin{eqnarray}
S^{(1)L}_g\left(k_1,k_2 , \frac{P_h}{z} \right)&& = \frac{i N_c}{2}
\bar{\Lambda}_{\alpha}(k_2 , \frac{P_h}{z}  )
{\cal S}_{+}(k_2 + q -\frac{P_h}{z})
\Lambda_{\beta} (k_1, \frac{P_h}{z} + k_1 - k_2 )
\left[ -\hat{g}_{t}^{\alpha \eta}(P_{h})\right]  
\nonumber\\
\times
&&\!\!\!
V_{\sigma \eta \beta}(k_2 - k_1 , -\frac{P_h}{z},  \frac{P_h}{z}+k_1 -k_2 )p^\sigma 
\frac{-i}{(\frac{P_h}{z} +k_1 -k_2)^2 + i\varepsilon}\ ,
\label{sLg}
\end{eqnarray}
where 
the relevant photon-quark-gluon
vertex functions $\bar{\Lambda}_{\alpha}(k_2 , P_h /z  )$ and 
$\Lambda_{\beta} (k_1, P_h /z + k_1 - k_2 )$
can be expressed using that of Fig.~\ref{fig3} by appropriate substitutions of the 
momenta, $\Lambda_\alpha (k_2,  P_h /z ) = \Gamma_\alpha (k_2 , k_2 +q-P_h /z )$.
$V_{\mu_1 \mu_2 \mu_3}(q_1 , q_2 , q_3)=(q_1 -q_2 )_{\mu_3}g_{\mu_1 \mu_2} +$ (cyclic permutation)
denotes the ordinary three-gluon vertex except for the color structure that has been used
as $f^{cab} t^a t^b= (iN_c /2 ) t^c$ to obtain the prefactor $iN_c /2$,
the structure 
$-\hat{g}_{t}^{\alpha \eta}(P_h )= -g^{\alpha \eta} 
+ (P_{h}^{\alpha} p^{\eta}+P_{h}^{\eta} p^{\alpha} )/(P_h\cdot p)$ 
projects the final-state gluon onto the twist-2 fragmentation 
process to produce the pion (see (\ref{Wmunu})), and 
${\cal S}_{+}(k)= 2\pi
\delta\left(k^2 \right) \theta (k^0 )\slash{k}$
is the cut propagator for the final-state hard quark. 
The mirror diagrams of Fig. 5 gives 
$S^{(1)R}_g(k_1,k_2,P_h/z)=\bar{S}^{(1)L}_g(k_2,k_1,P_h/z)$.  
Similarly to (\ref{deco}), $S^{(1)L}_g$ can be decomposed as
\beq
S^{(1)L}_g\left(k_1,k_2 , \frac{P_h}{z}\right)=
S^{(1\cdot)L}_g\left(k_1,k_2 , \frac{P_h}{z}\right)+
S^{(1\perp)L}_g\left(k_1,k_2 , \frac{P_h}{z}\right),
\label{decog}
\eeq
by using (\ref{psigma}). 
The three-gluon vertex coupling to the scalar-polarized coherent gluon 
in $S^{(1\cdot)L}_g(k_1,k_2,{P_h / z})$
can be 
disentangled using Ward identity similarly to (\ref{Phi}),
but this time we also obtain some additional ghost-like gauge terms.
It is indeed not difficult to show that those gauge terms drop in the final result
applying Ward identities further. Alternatively, one may employ
background field gauge\,\cite{Abbott}: 
In the Feynman gauge under the background coherent-gluon field,
the three-gluon vertex in (\ref{sLg}) is replaced as 
$V_{\sigma \eta \beta}(k_2 - k_1 , -P_h /z, P_h /z+k_1 -k_2 )
\rightarrow
V^{BG}_{\sigma \eta \beta}(k_2 - k_1 , - P_h /z, P_h /z+k_1 -k_2 )\equiv
V_{\sigma \eta \beta}(k_2 - k_1 , - P_h /z, P_h /z+k_1 -k_2 )
- (P_{h}/z)_\eta g_{\sigma \beta} -(P_h /z+k_1 -k_2 )_\beta g_{\sigma \eta}$,
and the resulting $S^{(1\cdot)L}_g(k_1,k_2,{P_h / z})$ obeys simple Ward identity
without unwanted gauge terms,
which is analogous to Fig.~\ref{fig:Ward}
for the quark-gluon vertex.
As a result, we get for 
$S^{(1\cdot)}_g(k_1,k_2,P_h/z)=
S^{(1\cdot)L}_g(k_1,k_2,P_h/z)+
S^{(1\cdot)R}_g(k_1,k_2,P_h/z)$
\begin{equation}
S^{(1\cdot)}_g\left(k_1,k_2, \frac{P_h}{z}\right)
=\frac{N_c}{2}\frac{k_{2\perp}^\sigma -k_{1\perp}^\sigma}{x_2 - x_1 - i\varepsilon}
 \left. \frac{\partial}{\partial r^{\sigma}_\perp} 
\bar{\Lambda}_{\alpha}(x_1 p , r)
{\cal S}_{+}(x_1 p+ q - r)
\Lambda_{\beta} (x_1 p , r)
\hat{g}_{t}^{\alpha \beta}(P_{h}) 
\right|_{r\rightarrow \frac{P_h}{z}} ,
\label{wardg2}
\end{equation}
similarly to (\ref{ward2}), up to the irrelevant higher order terms. 

The contribution $S^{(1\perp)L}_g(k_1,k_2,{P_h / z})$ in (\ref{decog})
can be also reduced similarly to (\ref{sL2}): We can 
use the decomposition of the product of the three-gluon vertex and the gluon propagator,
adjacent to
the fragmentation insertion $\hat{g}_{t}^{\alpha \eta}(P_h )$,
into the eikonal propagator term and the contact term as
\begin{eqnarray}
\hat{g}_{t}^{\alpha \eta}(P_h )&&\!\!\!\!
 V^{BG}_{\sigma \eta \beta}(x_2 p - x_1 p , -\frac{P_h}{z},  \frac{P_h}{z}+x_1 p -x_2  p )
\frac{-1}{\left(\frac{P_h}{z} +(x_1  -x_2 ) p\right)^2 + i\varepsilon} 
\nonumber\\
&&= \hat{g}_{t}^{\alpha \eta}(P_h ) 
\left(- \frac{P_{h \sigma}g_{\eta \beta}}{P_h \cdot p} \frac{1}{x_2 -x_1 - i\varepsilon} 
+g_{\sigma \eta}p_\beta 
\frac{z}{P_h \cdot p}\right)\ ,
\label{eikog}
\end{eqnarray}
for $\sigma = \perp$. This simple formula analogous to (\ref{eiko})
holds in the background field gauge mentioned above. 
Again, only the single pole terms 
eventually survive in the total contribution, $S^{(1\perp)}_g(k_1,k_2,{P_h / z})
=S^{(1\perp)L}_g(k_1,k_2,{P_h / z}) 
+ S^{(1\perp)R}_g(k_1 , k_2 , {P_h / z})$, and 
the result is given by the RHS of (\ref{perppart}) with the substitutions
necessary for ``translating'' (\ref{ward2}) into (\ref{wardg2}).
Combining this result with (\ref{wardg2}) and taking the 
similar steps as those in (\ref{total})-(\ref{wmunu}), we find that
$w_{\mu \nu}$ for the gluon fragmentation channel 
is given by (\ref{wmunu}) with 
the replacement
$1/N_c \rightarrow - N_c$ and 
$S^{(0)} (x p, P_h /z) \rightarrow S^{(0)}_g(x p, P_h /z) $, where
\begin{equation}
S^{(0)}_g(x p, \frac{P_h}{z} )=C_F \bar{\Lambda}_{\alpha}(x p , \frac{P_h}{z} )
{\cal S}_{+}(x p + q -\frac{P_h}{z})
\Lambda_{\beta} (x p , \frac{P_h}{z} )
[-\hat{g}_{t}^{\alpha \beta}(P_{h}) ]\ .
\label{s0g}
\end{equation}
This $S^{(0)}_g (x p, P_h /z)$  is the hard scattering function 
for the 2-to-2 partonic Born subprocess leading to the gluon fragmentation, 
and participates in the twist-2 contribution for the unpolarized SIDIS
as (\ref{wmunutw2}) with
$S^{(0)} (x p, P_h /z) \rightarrow S^{(0)}_g(x p, P_h /z)$.

Substituting 
(\ref{s0})-(\ref{wmunutw2}) and (\ref{s0g}) into 
(\ref{Wmunu}), we contract the result with the leptonic tensor for the unpolarized electron, 
$L_{\mu\nu}(\ell,\ell')=2(\ell_\mu \ell'_\nu+\ell_\nu \ell'_\mu )-g_{\mu\nu}Q^2$.
Working out the phase space factor corresponding to the differential elements,
$[d\omega]= dx_{bj}dQ^2 dz_f dq_T^2 d\phi$,
with the kinematical variables introduced above (\ref{twist3distr}) and 
the azimuthal angle $\phi$ for the observed final-state pion (see \,\cite{ekt06,EKT}),
we immediately obtain the formula 
for the single-spin-dependent cross section in SIDIS, $ep^\uparrow\to e\pi X$,
associated with
the soft-gluon mechanism induced by the twist-3 effects inside the nucleon, as
\begin{equation}
\frac{d\sigma^{\rm SGP}_{\mbox{\scriptsize tw-3}}}{[d\omega]}= 
\frac{\pi M_N}{2C_F}
\epsilon^{\sigma pnS_\perp} \sum_{j=q,g}{\cal C}_j \int \frac{dz}{z}\int \frac{dx}{x} D_j(z) 
\frac{\partial H_{jq}(x, z, q_T^2 )}{\partial (P_{h \perp}^\sigma /z)} G_F^q(x,x)\ ,
\label{tw3formula}
\end{equation}
where 
the sum over all quark and antiquark flavors $q=u, \bar{u}, d, \bar{d}, \cdots$ is implicit for 
the index $q$, and $G_F^q(x,x)$ denotes the ``soft-gluon-pole function'' from (\ref{twist3distr}) 
for the flavor $q$. The color factors are introduced as 
${\cal C}_q = -1/( 2N_c )$ and $1/(2N_c )$ for quark and antiquark flavors, 
respectively,~\footnote{The relative minus sign in ${\cal C}_q$ between quark and antiquark is 
due to that of the color charge between them, to which the coherent gluon couples. (See Erratum in pp.18-20.)} 
and ${\cal C}_g = N_c/2$. 
$H_{jq}(x, z, q_T^2)$ for $j=q$ and $g$ are, respectively, equal to
${\rm Tr}\left[ S^{(0)} (x p, P_h /z)\pslash \right]$ 
and 
${\rm Tr}\left[ S^{(0)}_g (x p, P_h /z)\pslash \right]$, 
up to the kinematical factor.   They are exactly the partonic hard scattering cross sections 
which participate in the twist-2 factorization formula of 
the unpolarized cross section for $ep \rightarrow e\pi X$ as
\begin{equation}
\frac{d\sigma^{\rm unpol}_{\mbox{\scriptsize tw-2}}}{[d\omega]}= 
\sum_{j=q,g}  \int \frac{dz}{z} \int \frac{dx}{x} 
D_j(z) H_{jq}(x, z, q_T^2 ) f_q(x)\ .
\label{tw2formula}
\end{equation}
Our results (\ref{tw3formula}) and (\ref{tw2formula}) represent
the SGP contribution as the response of 2-to-2 partonic Born subprocess
to the change of the transverse momentum carried by the ``fragmenting parton''. 
It is worth noting that our results (\ref{tw3formula}) and (\ref{tw2formula}) hold
in any Lorentz frame with $p_\perp =0$, as is seen from the above derivation.
In particular, we recall that 
the derivative with respect to the transverse momentum in (\ref{tw3formula}) was
introduced merely as a formal recipe via the relation (\ref{eq:relrel}), so that
one can freely move to any frame even with $P_{h \perp}=0$
after performing the derivative.

In the hadron frame mentioned above (\ref{twist3distr}),
$P_h^\sigma$ is parameterized as
$\sqrt{-P_{h\perp}^2}=z_fq_T$, $P_h^-=z_fQ/\sqrt{2}$ 
and $P_h^+=-P_{h\perp}^2/2P_h^-=z_fq_T^2/\sqrt{2}Q$, and thus
the derivative
on the Lorentz-scalar functions $H_{jq}(x, z, q_T^2 )$ 
with respect to $P_{h\perp}^\sigma$ can be performed
through $q_T$ that is indicated explicitly in their argument. 
Therefore the results 
(\ref{tw3formula}) and (\ref{tw2formula}) can 
be expressed as~\footnote{See Erratum in pp.18-20.} 
\begin{equation}
\frac{d\sigma^{\rm SGP}_{\mbox{\scriptsize tw-3}}}{[d\omega]}= 
\frac{\pi M_N}{C_F z_f^2}
\epsilon^{pn S_\perp P_{h \perp}} \left. \frac{\partial}{\partial q_T^2}
\frac{d\sigma^{\rm unpol}_{\mbox{\scriptsize tw-2}}}{[d\omega]}
\right|_{f_q(x)\rightarrow G_F^q(x,x),\ D_j(z) \rightarrow {\cal C}_j zD_j (z)}\ .
\label{sidis}
\end{equation}
The partonic Born cross section in (\ref{tw2formula})
was derived as\,\cite{Mendez78,MOS92,KN03}
\begin{equation}
H_{jq}(x, z, q_T^2 ) = {\alpha_{em}^2 \alpha_s e_q^2 \over 8\pi x_{bj}^2 S_{ep}^2 Q^2}
\sum_{k=1}^4 {\cal A}_k  
\widehat{\sigma}^{jq}_k
\delta\left( {q_T^2\over Q^2} -
\left( {1\over \xhat} -1\right)\left({1\over \zhat}-1\right)\right)\ ,
\label{hardc}
\end{equation}
where ${\cal A}_1 =1+\cosh^2\psi,
{\cal A}_2 =-2,
{\cal A}_3 =-\cos\phi\sinh 2\psi$, and
${\cal A}_4 =\cos 2\phi\sinh^2\psi$, with $\cosh\psi = 2x_{bj}S_{ep}/ Q^2 -1$,
parameterize different azimuthal dependence, 
and we introduced auxiliary variables $\xhat= x_{bj}/ x$ and $\zhat= z_f / z$.
$\widehat{\sigma}^{jq}_k$ are the functions of $\xhat, \zhat, q_T^2$, and $Q^2$,
whose explicit form can be found in Eqs.~(57) and (59) of \cite{ekt06}.
Therefore, the derivative $\partial /\partial q_T^2$ of (\ref{sidis}) can act on either 
$\widehat{\sigma}^{jq}_k$ or the delta-function 
$\delta\left( q_T^2 / Q^2 - ( 1/ \xhat -1)(1/ \zhat-1)\right)$, and the latter contribution
produces the ``derivative term'' proportional to $d G_F^q (x, x)/dx$ as well as the 
``non-derivative term'' with $G_F^q (x, x)$, after the partial integration with respect to $x$.
We get~\footnotemark[3]
\begin{eqnarray}
\lefteqn{
\frac{d\sigma^{\rm SGP}_{\mbox{\scriptsize tw-3}}}{dx_{bj}dQ^2 dz_f dq_T^2 d\phi}
= {\alpha_{em}^2 \alpha_s e_q^2 \over 8\pi x_{bj}^2 S_{ep}^2 Q^2}
\frac{\pi  M_N}{C_F z_f Q^2} 
\epsilon^{pnS_\perp P_{h\perp}}
\sum_{k=1}^4 {\cal A}_k \sum_{j=q,g}{\cal C}_j}
\nonumber\\
&&\;\;\;\;\;\;\times
\int \frac{dz}{z} \int \frac{dx}{x} D_j(z) 
\left\{  \frac{\xhat}{1-\zhat}
\widehat{\sigma}^{jq}_k  x\frac{dG_F^q (x, x)}{dx}
+\left[ 
\frac{1}{\zhat}Q^2
\frac{\partial \widehat{\sigma}^{jq}_k}{\partial q_T^2} 
-\frac{\xhat}{1-\zhat}\frac{\partial (\xhat \widehat{\sigma}^{jq}_k )}{\partial \xhat}
\right] G_F^q (x, x)\right\}
\nonumber\\
&&\;\;\;\;\;\;\times
\delta\left( {q_T^2\over Q^2} -
\left( {1\over \xhat} -1\right)\left({1\over \zhat}-1\right)\right) .
\label{sidisf}
\end{eqnarray}
Substituting the explicit formulae for $\widehat{\sigma}^{jq}_k$,
it is straightforward to see that this result completely coincides
with that
obtained recently in \cite{ekt06} by direct evaluation of each Feynman diagram in 
Figs.~\ref{fig1} and \ref{fig:gluon},
for all azimuthal dependence, where ${\cal A}_{1,2}$ give the same azimuthal dependence as
the Sivers effect\,\cite{Sivers} while ${\cal A}_3$ and ${\cal A}_4$ give additional terms beyond the
Sivers effect. 
Our result reveals 
that not only the derivative term of the
SGP cross section but the whole partonic hard scattering functions for the SGP contributions
are completely determined from $\widehat{\sigma}^{jq}_k$.

We emphasize that our result (\ref{sidis}), and its immediate consequence (\ref{sidisf}),
show extremely nontrivial structure behind the SGP contribution,
which would not be unveiled without using our new approach discussed above.
Indeed 
it is completely hopeless to infer from the complicated formulae (82) and (87) of \cite{ekt06}
for the non-derivative term 
as a function of $\xhat, \zhat, q_T^2$, and $Q^2$ 
that 
the non-derivative term can be expressed in terms of $\widehat{\sigma}^{jq}_k$
as in (\ref{sidisf}) with a form common to all channel ($j=q, g$; $k=1, \cdots, 4$).
We also note an important point which is most clearly indicated by the compact form (\ref{sidis}):
Apparently the cross section (\ref{hardc})
for the 2-to-2 partonic Born subprocess is gauge-independent, and so is 
its derivative 
with respect to a kinematical variable $q_T$. 
Combined with (\ref{sidis}), this fact guarantees
the gauge invariance of the hard-scattering function for the twist-3 SGP contribution.
In this connection we recall that a straightforward check of gauge invariance 
is hampered for the SGP contribution, in contrast to the case for the SFP and HP contributions,
because Ward identities are useless
to the coupling of the zero-momentum coherent gluon.

\begin{figure}[t!]
\begin{center}
\epsfig{figure=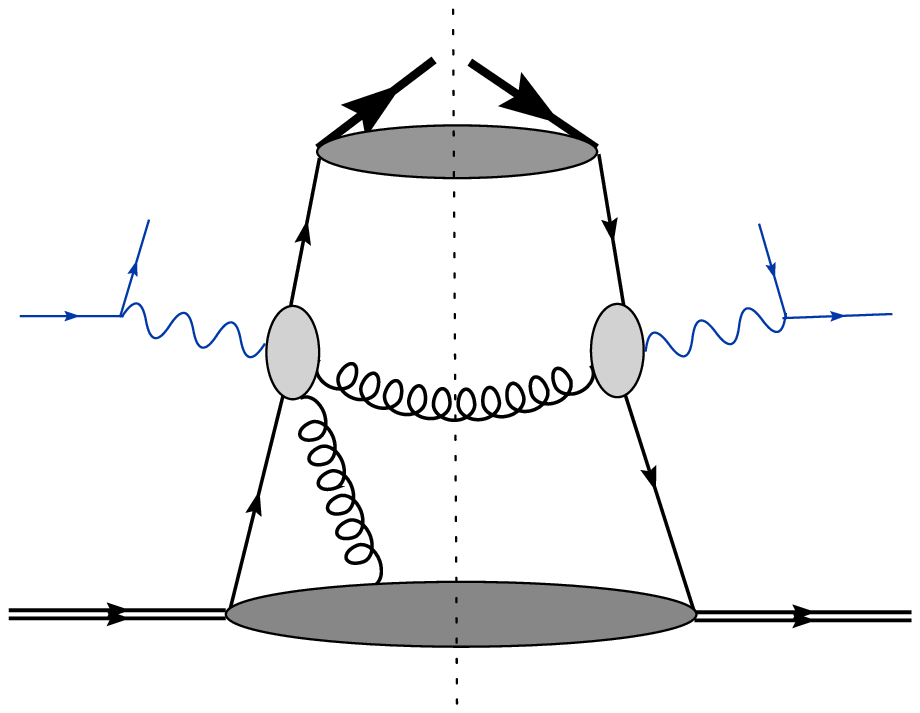,width=0.4\textwidth}
\hspace{0.6cm}
\epsfig{figure=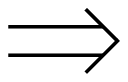,width=0.05\textwidth}
\hspace{0.6cm}
\epsfig{figure=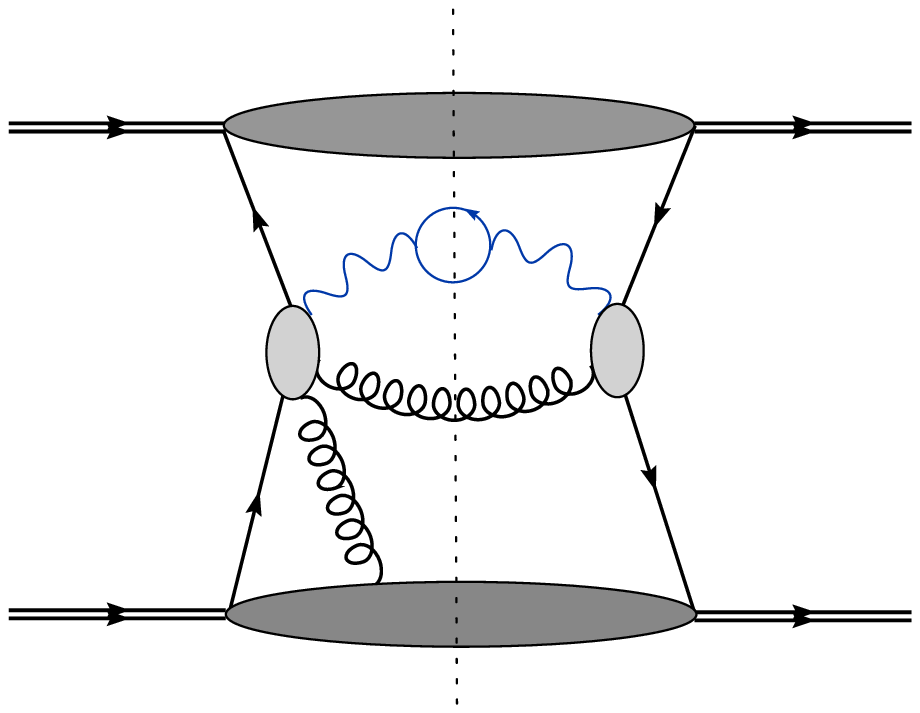,width=0.4\textwidth}
\end{center}
\caption{Diagrammatic representation of the crossing transformation of SIDIS into Drell-Yan process.
\label{fig:crossing}
}
\end{figure}

The above results (\ref{tw3formula}) and (\ref{tw2formula}) that lead 
to (\ref{sidis}), (\ref{sidisf})
have been derived using only the properties satisfied by the partonic subprocess  
in QCD perturbation theory. 
Therefore, 
by analytically continuing the external momenta,
the same formula represents the exact relations associated with other hard processes.
In fact, the corresponding formula for the Drell-Yan process can be obtained by
switching the final-state ``fragmenting parton'' into the initial-state
parton, and the initial-state photon with space-like momentum into the final-state 
photon with time-like momentum.
This crossing transformation, as illustrated in Fig.~\ref{fig:crossing},
is accomplished by the corresponding substitutions: 
$P_h \rightarrow - p'$, $1/z \rightarrow x'$, $D_q (z)\rightarrow f_{\bar{q}}(x')$, 
$D_g (z) \rightarrow f_g(x')$,
and $q^\mu \rightarrow -q^{\mu}$
in (\ref{tw3formula}) and (\ref{tw2formula}), where $f_{\bar{q}}(x')$ and $f_g(x')$ denote 
the twist-2 parton distributions for the unpolarized initial hadron with momentum $p'$,
and the new $q^{\mu}$ denotes the momentum of the virtual photon 
which is produced by the hard interaction of the 
partons and goes into the lepton pair $\ell^+ \ell^-$ with the invariant mass squared, 
$q^2 \equiv Q^2$, in the final state.
With this replacement, 
our ``master formula'' (\ref{tw3formula}) also describes the
single-spin-dependent cross section for the Drell-Yan process $p^\uparrow p\to \ell^+ \ell^- X$,
relating it to (\ref{tw2formula}) for the spin-averaged case $p p\to \ell^+ \ell^- X$.  
The partonic hard scattering cross sections are now expressed by 
the variables, $s=(p+p' )^2$, $\hat{s}=(xp +x' p')^2$, $\hat{t}= (xp -q)^2$, and
$\hat{u}=(x'p'-q)^2$, 
as (see \cite{JQVY06})
\begin{equation}
H_{jq} (\hat{s}, \hat{t}, \hat{u})= \frac{\alpha_{em}^2 \alpha_s e_q^2}{3\pi N_c s Q^2}
\widehat{\sigma}_{jq}^{DY}(\hat{s}, \hat{t}, \hat{u}) 
\delta \left( \hat{s}+\hat{t}+\hat{u}-Q^2 \right)\ ,
\label{hdy}
\end{equation}
with ($T_R = 1/2$)
\begin{equation}
\widehat{\sigma}_{\bar{q} q}^{DY}(\hat{s}, \hat{t}, \hat{u}) 
= 2C_F \left( \frac{\hat{u}}{\hat{t}}+\frac{\hat{t}}{\hat{u}} 
+ \frac{2Q^2 \hat{s}}{\hat{u}\hat{t}}
\right)\ , \;\;\;\;\;\;
\widehat{\sigma}_{g q}^{DY}(\hat{s}, \hat{t}, \hat{u}) 
= 2 T_R \left( \frac{\hat{s}}{-\hat{t}}+\frac{-\hat{t}}{\hat{s}} 
-\frac{2Q^2 \hat{u}}{\hat{s}\hat{t}}
\right)\ . 
\label{sigdy}
\end{equation}
Thus the derivative in (\ref{tw3formula}), which is now with respect 
to $-x' p_\perp^{\prime \sigma}$ instead of $P_{h\perp}^\sigma/z$,
\footnote{This corresponds to the fact that the SGP in Drell-Yan comes 
from the initial-state interaction,
while that in SIDIS is from the final-state interaction.}
can be performed through that for $\hat{u}$ (see the discussion below (\ref{tw2formula})).
After performing the derivative,
one can go over to a frame where the two colliding nucleons are collinear
along the $z$-axis, and
the produced virtual photon has large transverse momentum $q_\perp$, 
which is provided by 
the recoil from the hard parton going into the unobserved final state.  
We thus obtains, similarly to (\ref{sidis}) and (\ref{sidisf}),
\begin{eqnarray}
\lefteqn{
\frac{d\sigma^{\rm SGP, DY}_{\mbox{\scriptsize tw-3}}}{dQ^2 dy d^2 q_\perp}= 
\frac{\alpha_{em}^2 \alpha_s e_q^2}{3\pi N_c s Q^2}
\frac{\pi  M_N}{C_F} 
\epsilon^{pnS_\perp q_\perp}
\sum_{j=\bar{q}, g}{\cal C}_{\bar{j}}
\int \frac{dx'}{x'} \int \frac{dx}{x} \delta \left( \hat{s}+\hat{t}+\hat{u}-Q^2 \right)
f_j(x') 
}
\nonumber\\
&&\;\;\;\;\;\;\times
\left\{  \frac{\widehat{\sigma}^{DY}_{jq}}{-\hat{u}}  x\frac{dG_F^q (x, x)}{dx}
+\left[\frac{\widehat{\sigma}^{DY}_{jq}}{\hat{u}}  
-\frac{\partial \widehat{\sigma}^{DY}_{jq}}{\partial \hat{u}}  
- \frac{\hat{s}}{\hat{u}}\frac{\partial  \widehat{\sigma}^{DY}_{jq}}{\partial \hat{s}}  
-\frac{\hat{t}-Q^2}{\hat{u}}\frac{\partial  \widehat{\sigma}^{DY}_{jq}}{\partial \hat{t}}  
\right] G_F^q (x, x)\right\}\ ,
\label{dyf}
\end{eqnarray}
where $y$ is the rapidity of the virtual photon, and ${\cal C}_{\bar{g}}\equiv {\cal C}_{g}$.
This obeys exactly the same pattern as (\ref{sidisf}) for both derivative and non-derivative terms.
Substituting (\ref{sigdy}), this formula
completely coincides with the result of \cite{JQVY06} which
was obtained by direct evaluation of the Feynman diagrams.

We can also derive  
the single-spin-dependent cross section
for the direct-photon production, $p^\uparrow p\to \gamma X$, 
immediately:
We make  
the formal 
replacement $\alpha_{em}/(3 \pi Q^2 ) \rightarrow \delta(Q^2 )$
in (\ref{dyf}), corresponding to the real photon in the final state,
and define
$\widehat{\sigma}^{DP}_{jq}$ as the $Q^2 \rightarrow 0$ limit of
$\widehat{\sigma}^{DY}_{jq}$ of (\ref{sigdy}).
Because $\widehat{\sigma}^{DP}_{jq}$ are invariant 
under the scale transformation of the partonic variables $\hat{s}, \hat{t}$ and $\hat{u}$,  
we have 
$(\hat{u}\partial / \partial \hat{u}  
+\hat{s} \partial / \partial \hat{s}  
+ \hat{t} \partial  / \partial \hat{t} ) \widehat{\sigma}^{DP}_{jq}=0$.  
Accordingly, 
the single-spin-dependent cross section for the direct-photon (DP) production reads
\begin{eqnarray}
E_{q}\frac{d\sigma^{\rm SGP, DP}_{\mbox{\scriptsize tw-3}}}{d^3 q}&= &
\frac{\alpha_{em} \alpha_s e_q^2}{N_c s}
\frac{\pi  M_N}{C_F} 
\epsilon^{pnS_\perp q_\perp}
\sum_{j=\bar{q}, g}{\cal C}_{\bar{j}}
\int \frac{dx'}{x'} \int \frac{dx}{x} \delta \left( \hat{s}+\hat{t}+\hat{u} \right)
f_j(x') 
\nonumber\\
&&
\times
\frac{\widehat{\sigma}^{DP}_{jq}}{-\hat{u}}\left( x\frac{dG_F^q (x, x)}{dx} - G_F^q (x, x)
\right)\ ,
\label{dpf}
\end{eqnarray}
with $E_q = |\vec{q}|$.   Note the same twist-2 unpolarized hard cross section appears
both for derivative and
non-derivative terms, 
as the coefficient for
the combination, $x dG_F^q (x, x)/dx - G_F^q (x, x)$.
This remarkably compact result does not agree with that of \cite{QS91}\,\footnote{
This is because our result (\ref{dyf}) for the Drell-Yan process agrees with
that in \cite{JQVY06} which is different from the result in \cite{QS91} 
in the $Q^2\to 0$ limit.}, but
is reminiscent 
of the recent result for $p^\uparrow p \to\pi X$ \cite{KQVY06}, where
a similar compact formula was obtained for the SGP cross section.  
One can extend the present approach to other processes such as $p^\uparrow p \to\pi X$ and show
that the compact result found in \cite{KQVY06} is also a consequence of 
the simplification
due to the scale invariance of the 2-to-2 Born subprocess among massless partons,
similarly to the 
present case (\ref{dpf})\,\cite{KT07}.

To summarize, we have studied the SGP contribution in the 
twist-3 mechanism for the SSA. 
We have developed a new approach that allows
systematic reduction of the coupling of the soft coherent-gluon and the associated 
pole contribution,
using Ward identities and decomposition identities for the interacting parton propagator, 
and derived 
the master formula which gives the twist-3 SGP contributions to the SSA
entirely in terms of the knowledge of the twist-2 factorization formula 
for the unpolarized cross
section. We find that this novel result is also useful for establishing the gauge invariance of 
the hard-scattering function for the SGP contribution.
So far this master formula was derived
for DY process, direct $\gamma$ production and SIDIS.
Since our diagrammatic manipulation technique 
uses only elementary identities which hold for any diagram with the gluon insertion to cause the SGP, 
it is applicable to other processes such as $p^\uparrow p \to\pi X$ and
$pp\to \Lambda^\uparrow X$ etc, and 
may lead to a similar relation between the twist-3 SGP contribution and the twist-2 
cross section, which allows us to
reveal the corresponding gauge-invarinant structure.

\section*{Acknowledgements}
We thank Feng Yuan for useful comments.  
The work of K.T. was 
supported by the Grant-in-Aid for Scientific Research No. C-16540266. 


\newpage

\begin{center}
{\large\bf Erratum to: ``Master formula for twist-3 
soft-gluon-pole mechanism to single transverse-spin asymmetry''}\\
{\large\bf [Phys. Lett. B 646 (2007) 232 (arXiv:hep-ph/0612117)]}
\end{center}
\begin{center}
{\normalsize
Yuji Koike\ $^{\rm a}$,
Kazuhiro Tanaka\ $^{\rm b}$}
\end{center}

\begin{center}
{\footnotesize
$^{\rm a}${\it Department of Physics, Niigata University, Japan}\\
$^{\rm c}${\it Department of Physics, Juntendo University, Japan}
}
\end{center}


%
%
\baselineskip 18pt

\bigskip

There was an error in Eqs.~(24) and (26) of our Letter~[1].
These equations were derived 
by performing the derivative in Eq.~(22) with respect to the transverse component 
of the final-pion momentum, $P_{h\perp}^\sigma
= (P_{h\perp}^+, P_{h\perp}^- , {\mathbf{P}}_{h\perp})= 
(0, 0, {\mathbf{P}}_{h\perp})$.
The derivative for
the variation of the magnitude of ${\mathbf{P}}_{h\perp}$ 
was
performed through $q_T$ in the hadron frame, as explained above Eq.~(24),
but 
that for the variation of the direction of ${\mathbf{P}}_{h\perp}$
also proves to yield nonzero contribution.
The latter can be performed through 
$\phi$ associated with ${\cal A}_3$, ${\cal A}_4$ in Eq.~(25), where $\phi$ is defined as
the azimuthal angle of the initial-lepton's momentum 
measured from the axis along ${\mathbf{P}}_{h\perp}$ in the hadron frame
(see Refs.~[6,15] in [1]).
The correct equations are 
\setcounter{equation}{23}
\begin{equation}
\frac{d\sigma^{\rm SGP}_{\mbox{\scriptsize tw-3}}}{[d\omega]}= 
\frac{\pi M_N}{C_F z_f^2}
\left(
{\epsilon^{pn S_\perp P_{h\perp}}}
\left. \frac{\partial}{\partial q_T^2}
-\left(P_{h\perp}\cdot S_\perp\right)
\frac{1}{2 q_T^2}\frac{\partial}{\partial \phi}
\right)
\frac{d\sigma^{\rm unpol}_{\mbox{\scriptsize tw-2}}}{[d\omega]}
\right|_{f_q(x)\rightarrow G_F^q(x,x),\ D_j(z) \rightarrow {\cal C}_j zD_j (z)}\ ,
\end{equation}
and
\setcounter{equation}{25}
\begin{eqnarray}
\lefteqn{
\frac{d\sigma^{\rm SGP}_{\mbox{\scriptsize tw-3}}}{dx_{bj}dQ^2 dz_f dq_T^2 d\phi}
= {\alpha_{em}^2 \alpha_s e_q^2 \over 8\pi x_{bj}^2 S_{ep}^2 Q^2}
\frac{\pi  M_N}{C_F z_f Q^2} \sum_{j=q,g}{\cal C}_j
\left\{
\epsilon^{pnS_\perp P_{h\perp}}
\sum_{k=1}^4 {\cal A}_k 
\right.}
\nonumber\\
&&\;\;\;\;\;\;\times
\int \frac{dz}{z} \int \frac{dx}{x} D_j(z) 
\left(  \frac{\xhat}{1-\zhat}
\widehat{\sigma}^{jq}_k  x\frac{dG_F^q (x, x)}{dx}
+\left[ 
\frac{1}{\zhat}Q^2
\frac{\partial \widehat{\sigma}^{jq}_k}{\partial q_T^2} 
-\frac{\xhat}{1-\zhat}\frac{\partial (\xhat \widehat{\sigma}^{jq}_k )}{\partial \xhat}
\right] G_F^q (x, x)\right)
\nonumber\\
&&\;\;\;\;\;\;\times
\delta\left( {q_T^2\over Q^2} -
\left( {1\over \xhat} -1\right)\left({1\over \zhat}-1\right)\right) 
\\
&&
+ \frac{Q^2}{q_T^2}\! \left(P_{h\perp}\cdot S_{\perp}\right)\!\!
\int \frac{dz}{z} \int \frac{dx}{x}
D_j(z)\! \left[ {\cal A}_8 \frac{\widehat{\sigma}^{jq}_3}{2\zhat}
+ {\cal A}_9 \frac{\widehat{\sigma}^{jq}_4}{\zhat}\right]\! G_F^q (x, x)
\left. \! \delta \! \left( {q_T^2\over Q^2} -
\left( {1\over \xhat} -1\right)\left({1\over \zhat}-1\right)\right) \right\} .
\nonumber
\end{eqnarray}
Here the new term with the derivative with respect to $\phi$ in the former formula
yields the last line in the latter, with 
${\cal A}_8 =-\sin\phi\sinh2\psi$ and ${\cal A}_9 =\sin 2\phi\sinh^2\psi$,
which parameterize the novel azimuthal dependence (see  Refs.~[16,18] in [1]).
We note $\epsilon^{pnS_\perp P_{h\perp}}=z_f q_T\sin\Phi_S$ and 
$\left(P_{h\perp}\cdot S_{\perp}\right)=-z_f q_T\cos\Phi_S$, with $\Phi_S$
that denotes the azimuthal angle for the initial-proton's transverse spin vector $S_{\perp}$
measured from the axis along ${\mathbf{P}}_{h\perp}$ in the hadron frame.
{}From the above result (26), it is clear that the twist-3 single-spin-dependent cross section for SIDIS
consists of five independent structures having different dependences on the two azimuthal angles, 
$\phi$ and $\Phi_S$.
We remind that the similar five
structures also appear in the SIDIS single-spin-dependent cross section obtained by using
the transverse-momentum-dependent (TMD) parton distribution functions as shown in [2].
Thus, with the new terms in (26), 
the two approaches,
our twist-3 mechanism and the TMD one,
yield consistent azimuthal components for SSA.

\smallskip

It is instructive to consider the integration of the above formula (26) 
over the transverse momentum of the outgoing pion, ${\mathbf{P}}_{h\perp}$.
For this purpose
we introduce the azimuthal angles $\phi_h$ and $\phi_S$ for the hadron plane 
and 
the proton's transverse spin vector, respectively, as measured 
from the {\it lepton plane}.  
They are connected to our $\phi$ and $\Phi_S$ as
$\phi_h=\phi$ and $\phi_S=\phi-\Phi_S$.  Then the five different azimuthal structures in (26)
can be recast into those proportional to $\sin(\phi_h-\phi_S)$,
$\sin(\phi_h+\phi_S)$,
$\sin(2\phi_h-\phi_S)$,
$\sin(3\phi_h-\phi_S)$
and $\sin \phi_S$.  
Integrating (26) over the outgoing-pion's angle $\phi_h$ with $\phi_S$ kept fixed,
the contributions associated with the first four azimuthal structures vanish, leaving 
only the term proportional to $\sin \phi_S$.  However, further integration 
over $q_T^2$ ($=|{\mathbf{P}}_{h\perp}|^2/z_f^2$)
kills this term as well, as can be easily checked using
the above form (24); the $q_T^2$-integrals of two terms in (24) cancel out.
This result is consistent with the fact that there is no SSA in the totally inclusive DIS.

\smallskip

The results for the Drell-Yan and direct-photon production in the Letter~[1], Eqs.~(29) and (30),
which were also derived from Eq.~(22), 
remain unchanged, because those results were not concerned with the angular dependence
relevant to the above discussion.

\smallskip

All equations involving the quark-gluon correlation function
$G_F^{q}(x,x)$ for the flavor $q$
in the Letter were presented assuming that 
$G_F^{q}(x,y)$ for the antiquark flavors was given by Eq.~(1) with
the quark fields replaced by their charge-conjugate fields. 
However, this definition appears not to be completely consistent 
with charge-conjugation invariance. The appropriate definition of 
$G_F^{q}(x,y)$ for the antiquark flavors is given by Eq.~(1) with the quark-gluon 
three-body nonlocal operator replaced by its charge-conjugate nonlocal operator,
and in this case all equations involving 
$G_F^{q}(x,x)$, as well as the above formulae
(24) and (26), should read with 
${\cal C}_q = -1/( 2N_c )$ for both quark and antiquark flavors.

\smallskip

The conclusions made in the Letter are unaffected by all the above corrections.
%
\vspace{-0.11cm}
\section*{Acknowledgements}


The authors are grateful to M. Diehl, F. Yuan and W. Vogelsang for useful discussions.

\vspace{0.5cm}

\noindent
{\bf References}

\vspace{0.2cm}


\noindent
[1] Y. Koike and K. Tanaka, Phys. Lett. {\bf B646} (2007) 232 [arXiv:hep-ph/0612117].

\noindent
[2] A. Bacchetta, M. Diehl, K. Goeke, A. Metz, P.J. Mulders and M. Schlegal,
JHEP {\bf 02}\\ 
\hspace*{0.4cm} (2007) 093. 

\end{document}